# A Method to Improve the Resolution of the Acoustic Microscopy

I.V.Minin[1,*], Q. Tang[2], S. Bhuyan[3], J. Hu[2] and O.V.Minin[1]
[1]Siberian State University of Geosystem and Technologies, Novosibirsk, Russia
[2]Nanjing University of Aeronautics and Astronautics, Nanjing, China
[3]Siksha 'O' Anusandhan University, Bhubaneswar, India

*email: prof.minin@gmail.com

Abstract: In this report, we demonstrate a new principle to improve the resolution of the acoustic microscopy, which is based on the sub-wavelength focusing of acoustic wave passing through an acoustically transparent mesoscale particle. In the principle, the width of the acoustic focal area can be less than one wavelength. The sub-wavelength focusing effect is verified by the FEM simulation.

It is well known that the idea of the acoustic microscopy was proposed by S. Ya. Sokolov in 1936, to illustrate opaque media at the frequency of 3 GHz [1, 2]. The first practical ultrasonic absorption microscopy which worked at 12 MHz, was constructed at University of Illinois in 1959 [3]. Line-by-line sounding of a specimen by a focused ultrasonic beam was proposed by C. Quate and R. Lemons [4] in 1974, and they made it possible to obtain spatial distributions of local acoustic characteristics of a specimen and reconstruct ultrasonic images with high resolution.

Acoustic microscopy is an analogue of optical microscopy. The difference between them is that in light microscopy, optical contrasts of the studied structures are formed by reflection, scattering, and absorption of light, while in acoustic microscopy, the contrasts are formed by changes in the elasticity, density, and acoustic damping of the tested matters. Acoustic waves, actually mechanical elastic strain waves in a medium, are more sensitive to a change (break) in the acoustical impedance of the medium, which makes it possible to use them for detection and visualization of objects with surface and subsurface discontinuities and inclusions, and for measuring local physico-mechanical properties of different kinds of materials, including nontransparent ones. By now, in many scientific centers of the world, different original methods of the acoustic scanning microscopy have been implemented [5, 6].

The imaging quality of acoustic microscopy is highly determined by the characteristics of acoustic transducer, such as the resonant frequency, bandwidth [7], and focusing device. The bandwidth determines the axial resolution, while the acoustic radiation from the transducer determines the lateral resolution.

One major component in the performance of an acoustic microscope is the transducer-lens system (Fig.1a). Typically, a piezoelectric material (a buffer sapphire rod) attached to one end is used to excite acoustic wave, and a spherical depression (lens) formed in the other end is used to focus acoustic wave on the sample according to the Snell's law. In fact, due to large difference of the sound velocity between sapphire and water (a factor of 7.4), the focused beam suffers negligible spherical aberration and converges to a near diffraction-limited spot [8]. The object to be examined is placed at or near the focus. It is mechanically scanned line by line in a raster pattern. Acoustic power reflected by the object is collected and re-collimated by the lens and detected (in a phase-sensitive way) by the transducer.

Resolution is determined by the diameter of the focal spot, and the spot diameter is determined by the wavelength of the acoustic radiation in the coupling medium [9, 10]. Increasing the operating frequency of the instrument will improve the resolution by decreasing the acoustic wavelength. However, the maximum operating frequency is limited. The attenuation in the coupling fluid has a square law dependence on frequency extending from several kHz to about over 6 GHz (The acoustic attenuation depends on the molecular weight of the sample. Speed of sound is closely related to the density and elastic bulk modulus of the specimen). Thus, attenuation is a decisive parameter for the choice of the coupling fluid [11].

The choice of the operating frequency of emission is always a compromise between the size of the inhomogeneity to be visualized and the reached penetration depth in this case. Although increasing the frequency of emission can improve the resolution and make it possible to visualize micro objects, it also makes the view field narrower and, consequently, decreases the probability of finding a flaw and increases the requirements for the quality of surface treatment.

Increasing of resolution also can be obtained by heating the water (or a liquid of lower acoustic attenuation than water) to a higher temperature (because the acoustic attenuation in water decreases with temperature increase) and/or by using a lens of smaller focal length.

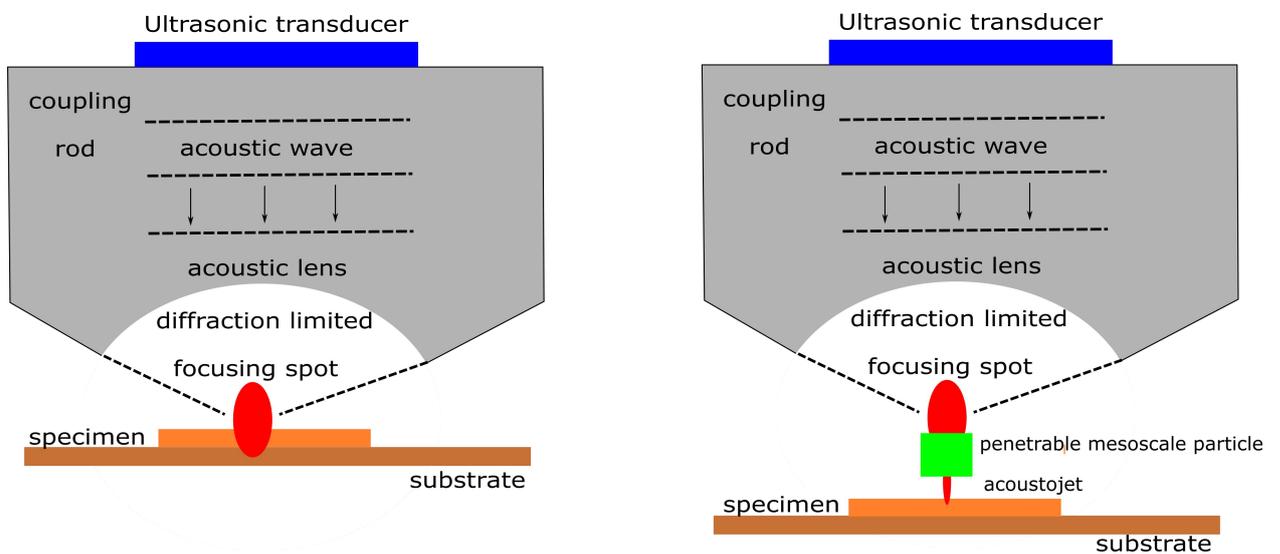

Fig.1. Scheme of an acoustical microscopy: a) classical and b) offered [28].

The principle of acoustic lens design is described as follows. Historically, the possibility of making an acoustical lens, analogous in function and operation to an optical lens, probably were first described by W. E. Kock and F. K. Harvey who were working on the refraction effects in conjunction with microwave transmission (Fig.2a) [12,13].

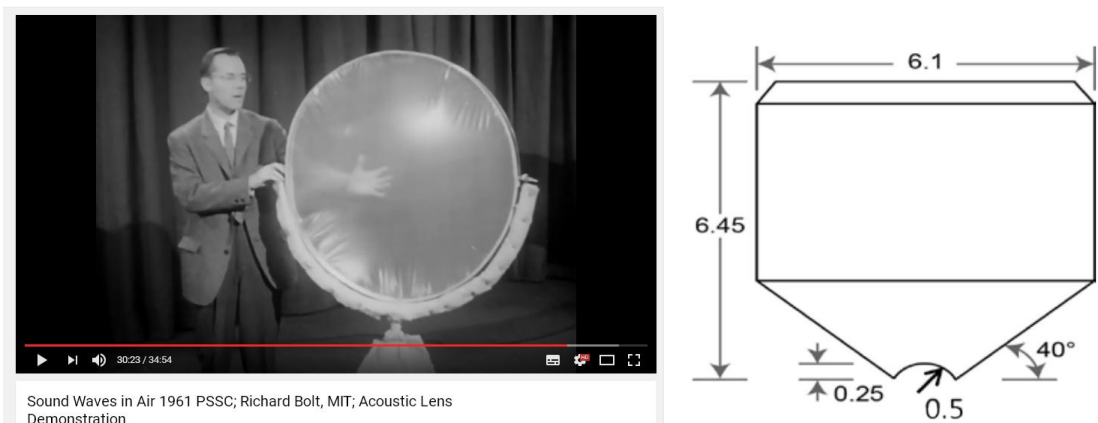

Fig.2. a) Demonstration of acoustic lens. Source: https://www.youtube.com/

b) Diagram of 400 MHz Acoustic Lens (Olympus, Model: AL4M631). The dimension unit in the figure is cm.

In an acoustic microscopy, as it is mentioned above, most of the simpler acoustic lens designs are assumed that the lens is placed close enough to the transducer and acts as a plane-wave source [14] (Fig.2b). The F-number (ratio of focal distance to lens diameter) of the acoustic lens need to be chosen to obtain a certain spot size for surface imaging applications [15].

Propagation of acoustic waves is controlled by the mass density and the bulk modulus of the acoustic medium through acoustic wave equation:

$$\nabla^2 P - \frac{\rho}{B}\frac{\partial^2 P}{\partial t^2} = 0,$$

where $P$ is the pressure, and $\rho$, $B$ are the mass density and bulk modulus of the acoustic medium, respectively. Physically, the mass density is defined as mass per unit volume, and the bulk modulus reflects the medium's resistance to external uniform compression. These two parameters are analogous to the electromagnetic (EM) parameters, permittivity $\varepsilon$ and permeability $\mu$, as can be seen in the following expression of the refractive index $n$ and the impedance $Z$.

$$n = \sqrt{\varepsilon\mu}, \cdot\cdot Z = \sqrt{\mu/\varepsilon} \cdot\cdot (EM),$$
$$n = \sqrt{\rho/B}, \cdot\cdot Z = \sqrt{\rho B} \cdot\cdot (Acoustic)$$

where the mass density and the bulk modulus are always positive in conventional media.

During the past few decades, persistent efforts have been made to go beyond diffraction limit of resolution, for example through the use of acoustical metamaterials. The concept of acoustic metamaterials and demonstration of a cylindrical acoustic lens that can separate features much smaller than the wavelength of acoustic waves were reported in [16], where the authors provide a slightly difference on the original optical hyperlens by exploiting the differences between optical and acoustic waves. Now it is possible to design acoustic lenses for sub-diffraction imaging [16-20]. The modern review of artificially structured acoustic metamaterials is given in [21].

It should be noted that the problem of homogenization of acoustic crystals with small (wavelength scaled) dimensions is not trivial, and we need to answer the following question: What is the minimum size of a cluster at which its properties can be described by effective values of its acoustical parameters?

The photonic nanojet effect (PNJ), well known in optical band now [22], has motivated us to propose the concept of acoustic jets called "*acoustojets*" [23,24]. It has been theoretically predicted for the first time [23,24] that the existence of acoustic analogue of PNJ phenomenon is possible, providing an opportunity for subwavelength localization of acoustic field in shadow area of an arbitrary 3D penetrable mesoscale particle. It is important to note that there is a principle difference between optical and acoustical materials with the existence of a shear sound speed, as acoustic materials are anisotropic due to two sound speeds [25]. For example, in [26] it was shown that as a result of the appearance of two independently propagating wave beams, two focuses appear in the liquid medium. One focus is due to the propagation of the longitudinal wave inside the lens, and another is caused by the transverse wave. Recently, the phenomenon of acoustojet formation was experimentally verified by a spherical Rexolite particle immersed in water [27].

The concept of acoustic microscopy with subwavelength resolution [28] is shown in the Fig.1b, and the physical principle is similar to Ref. [29] where we demonstrate for the first time that a mesoscale dielectric cube can be exploited as a novel resolution-enhancer by simply placing it at the focused imaging point of a continuous wave THz imaging system. It has been shown that by using the enhancer at 125 GHz in THz imaging system, we may obtained a diffraction-limited focusing spot corresponding to 275 GHz, a 2.2 times higher frequency. It seems to suggest that in sizes comparable to the wavelength, it is possible to focus with a cuboid particle lens due to diffraction effects, which could overcome the refractive effects of a conventional lens.

The case of a Rexolite sphere scattered by a traveling plane wave is chosen as an example to illustrate this possibility in acoustics [25, 27]. The mass density and longitudinal and shear speeds of sound in the Rexolite sphere are $\rho_1$ =1049 kgm$^{-3}$, $C_L$ =2337 ms$^{-1}$ and $C_S$ =1157 ms$^{-1}$, respectively. The host medium is water with density $\rho_0$ =998 kgm$^{-3}$ and speed of sound $C_0$ =1493 ms$^{-1}$. The simulations were made based on the finite element method which was implemented in COMSOL Multiphysics (Comsol, Inc., Burlington, MA) with a mesh having 10 points per wavelength.

Fig.3a shows the normalized intensity distribution of the acoustojet [23] for the sphere along propagation axis (z-axis), which is obtained by our measurement and FEM simulation. The radius of sphere is 4.14$\lambda$, where $\lambda$ (=$C/f$) is the wavelength [27]. The jet length $L_j$ is defined to be 2.12$\lambda$. The transverse plot is shown in Fig.3b. The full width at half maximum (FWHM) was 0.5$\lambda$.

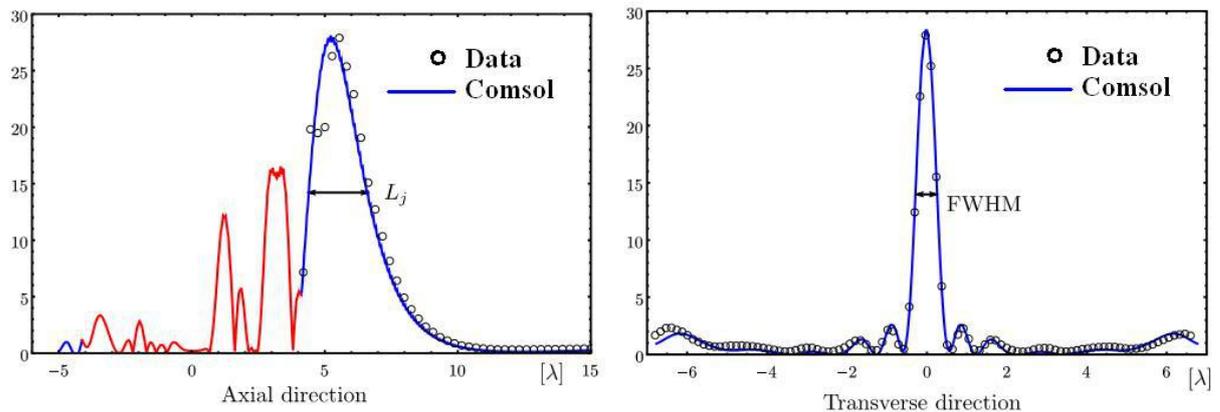

Fig.3. (a) Axial and (b) transverse normalized peak intensities of the superfocused beam by a Rexolite spherical particle lens with radius a =4.14$\lambda$ in water [27].

Conclusions
Our experimental and computational work indicates that a solid spherical particle lens can superfocus ultrasound beams. The FWHM of the focused beam for described example of penetrable particle-lens is $\lambda/2$ with an intensity gain of 14 dB. The concept of improved acoustic microscopy [28] is offered.